\documentstyle[aps,epsfig,preprint]{revtex}

\begin{document}
\tightenlines
\renewcommand{\thefootnote}{\fnsymbol{footnote}}
\title{Non-volatile molecular memory elements based on ambipolar nanotube
field effect transistors}

\author{M. Radosavljevi\'{c}\footnote{Present address: IBM TJ Watson
Research Center, Yorktown Heights, NY 10598}, M. Freitag, K.~V.
Thadani, and A.~T. Johnson\footnote{Corresponding author:
cjohnson@physics.upenn.edu}}
\address{Department of Physics and Astronomy and Laboratory for Research
on the\\Structure of Matter, University of Pennsylvania, Philadelphia, PA
19104, USA}
\date{\today}
\maketitle

\begin{abstract}
We have fabricated air-stable n-type, ambipolar carbon nanotube field
effect transistors (CNFETs), and used them in nanoscale memory cells.
N-type transistors are achieved by annealing of nanotubes in hydrogen gas
and contacting them by cobalt electrodes. Scanning gate microscopy reveals
that the bulk response of these devices is similar to gold-contacted
p-CNFETs, confirming that Schottky barrier formation at the contact
interface determines accessibility of electron and hole transport regimes.
The transfer characteristics and Coulomb Blockade (CB) spectroscopy in
ambipolar devices show strongly enhanced gate coupling, most likely due to
reduction of defect density at the silicon/silicon-dioxide interface
during hydrogen anneal. The CB data in the ``on''-state indicates that
these CNFETs are nearly ballistic conductors at high electrostatic doping.
Due to their nanoscale capacitance, CNFETs are extremely sensitive to
presence of individual charge around the channel. We demonstrate that this
property can be harnessed to construct data storage elements that operate
at the few-electron level.
\end{abstract}

\clearpage

Since the first demonstration of molecular carbon nanotube field
effect transistor (CNFET)\cite{tans_fet,martel_fet} just four
years ago, exploration of the potential of this system for use in
functional electronic devices has proceeded at a rapid pace.
Typical transistor circuits are p-type i.e., holes are majority
carriers, an effect originally ascribed to exposure to ambient
gas\cite{collins_oxygen}. Recently, scanning gate microscopy
(SGM)\cite{freitag_apl} as well as transport
experiments\cite{martel_prl,derycke} have shown that the charge
transport is influenced by formation of Schottky barriers at the
nanotube-electrode contact. More importantly, Derycke et
al.\cite{derycke,derycke_apl} can adjust the Schottky barrier
height using a combination of annealing and oxygen exposure, which
allows them to continuously tune from p-type to ambipolar to
n-type, i.e. electron transport. These advances have led to
integration of CNFETs into the first molecular logic
gates.\cite{derycke,bachtold_logic,liu_logic} To date however, all
n-CNFETs have been enclosed for protection from oxygen: either in
a vacuum chamber\cite{bock_Kdop} or by a silicon-dioxide
passivation layer\cite{martel_prl} which may not be suitable for
applications such as in sensor technologies.

In this Letter we demonstrate the first air stable n-type CNFET in
the more traditional open geometry. SGM data reveals that bulk
response of n-CNFETs is complementary to p-type CNFETs with gold
electrodes, verifying that carrier type is a function of the
barrier at the contact interface. Favorable Fermi level alignment
enables observation of ambipolar transport in these
devices.\cite{martel_prl} Our CNFETs show drastic increase in gate
coupling, which we attribute to reduction trap density at
Si/SiO$_2$ during the hydrogen anneal. For large negative gate
potential (hole transport regime) CNFET forms a single quantum dot
at low temperatures, suggesting that transport becomes ballistic.
Finally we exploit a hysteresis phenomenon to create molecular
memory cell based on a single CNFET that is stable on the scale of
days at room temperature.

CNFET devices are fabricated from nanotubes grown directly on the
substrate using chemical vapor
deposition.\cite{dai_nature,hafner_nature} Chrome/gold alignment
marks are patterned on degenerately doped p$^{++}$ Si wafers
capped with 225nm of thermally grown SiO$_2$. Next, catalyst
consisting of ethanol solution of fused alumina and ferric
nitrate\cite{snyder} is randomly dispersed on the substrate.
Nanotubes are grown by catalytic decomposition of ethylene gas at
$800^{\circ}$C.\cite{hafner_nature} While ethylene (2sccm) is only
streamed at setpoint temperature, hydrogen (400sccm) and argon
(600sccm) flow through the furnace during heating and cooling
cycles. Hydrogen has a two-fold role: first, it protects nanotubes
from burning by reacting with residual oxygen in the environment.
Second, it prevents degradation of the substrate and keeps the
trap density at Si/SiO$_2$ interface to a minimum.\cite{wolf} We
perform hydrogen anneal prior to contact fabrication in order not
to influence the Schottky barrier at the CNFET-electrode
interface. Following growth, nanotubes are located with respect to
alignment marks, and source and drain connections are fabricated
by electron beam lithography and liftoff.

Figure 1 shows current-gate voltage ($I-V_g$) transfer
characteristics for devices grown during the same run, and
fabricated with either chrome/gold (Cr/Au) or cobalt (Co)
electrodes. A distinct difference is always observed between two
sets of devices: Cr/Au-contacted CNFETs are intrinsically p-type,
while Co connection yields n-type $I-V_g$ behavior (Fig.~1(a)).
N-type conduction is stable under ambient conditions and observed
in twelve CNFETs grown in three different runs. To further examine
this issue we perform scanning gate microscopy
(SGM)\cite{tans_bach_sgm,freitag_ropes} on both p- and n-CNFETs
(Fig.~1(b)-(e)). The conductance of p-CNFETs is suppressed at
localized scattering sites within the nanotube by a positive SGM
tip voltage, $V_t$ (Fig.~1(c)). In contrast, a negative $V_t$ only
has an effect at the nanotube-metal junction, where it enhances
device conductance by increasing the transparency of the reverse
biased Schottky barrier at the source electrode.\cite{freitag_apl}
SGM reveals that the response of n-type, Co-contacted transistors
to local gating is complementary to that of p-CNFETs: Fig.~1(e)
shows that scattering sites along the nanotube are imaged for $V_t
< 0$. The Schottky barrier at the nanotube-cobalt interface is
weaker than in the case of Cr/Au electrodes (data not shown). Our
findings demonstrate the possibility of open-geometry air stable
n-CNFET, and support previous results\cite{martel_prl} that device
switching is influenced by the precise nature of the
nanotube-metal contact. Moreover, these data add to the mounting
evidence\cite{derycke} that suggests that hole density induced by
charge transfer between tube and adsorbed gases in the ambient is
not large enough to dominate device behavior under all
circumstances.

The observed p- to n-type conversion must be understood in terms
of precise control of the Schottky barrier at the device interface
depending on the contact scheme\cite{derycke}. Work function
difference alone can not account from the observed shift since
those of chrome (4.5eV) and cobalt (5.1eV) are both higher than
the nanotube workfunction (4.5eV).\cite{crc} In addition,
previously fabricated Co-CNFET devices are found to be
p-type.\cite{martel_iedm}  Some differences between two types of
devices are evident: First, contacts between metallic nanotubes
and cobalt have near-perfect transparency (as high as 98\%), while
Cr/Au contacts have transmission below 80\%.\cite{marko_unpub}
Second, they behave differently in the electron accumulation
regime. Cr/Au contacts lead to a sizeable Schottky
barrier\cite{freitag_apl,martel_prl} for electron injection at the
contacts. This occasionally leads to production of a nanoscale
quantum dot at the end of the nanotube.\cite{mceuen_pndot} We do
not observe this effect in n-type devices, suggesting a lower
barrier height for the cobalt-CNFET system, consistent with SGM
data. It is clear that two contact schemes are not equally
affected by processing and environmental conditions. This presents
an opportunity to differentiate the contributions of different
factors to the Schottky barrier height in nanotube systems, which
should allow individual control over device properties.

Figure 2 shows $I-V_g$ characteristics at fixed bias ($V_{ds}$)
for an $L=800$nm long, Co-contacted CNFET. At $V_g$=0, electrons
are the majority carriers, and the conductance decreases as the
gate voltage is made negative. Surprisingly, at more negative gate
voltage, the conductance increases again, a signature of room
temperature inversion to hole conduction in the device. Ambipolar
transport has been observed starting from a traditional p-CNFET in
our group and others.\cite{martel_prl,marko_unpub,mceuen_pndot} In
our samples, the resistance in inversion is typically five times
larger than in accumulation. Since device conductance, which is
less influenced by bulk scattering (see Coulomb blockade data
below) than tunnelling through the Schottky barrier at the source
electrode\cite{freitag_apl}, is similar in two transport regimes
so are the barriers for the two carrier types. Cobalt-contacted
CNFETs also show excellent switching characteristics for both
electrons and holes. From Fig.~2(b) we extract transconductances
of $2\mu$A/V and $3\mu$A/V at $V = 1$V for holes and electrons,
respectively, an order of magnitude better than previously
reported in solid-state devices. Using d$I/$d$V_g = \mu C_g
V/L^2$\cite{sze}, we infer a high carrier mobility on the order of
$\mu = 400$cm$^2$/V-s even at large 1V source-drain potential. As
shown below, the transport in CNFETs is not limited by scattering
in the nanotube bulk, and so the computed mobility is an
``effective'' quantity that characterizes device performance and
only a lower bound on the intrinsic mobility of the semiconducting
nanotubes. Lastly, our devices exhibit voltage gain ($\Delta
V_{ds}/\Delta V_g$, at $P = 0.5\mu$W; data not shown) between 2
and 3 in either regime, surpassed only by devices with
electrolyte\cite{kruger} and very thin aluminum oxide dielectric
layers.\cite{bachtold_logic} Voltage gains larger than 1 are
required in order to prevent signal degradation through multiple
logic elements. High gain in accumulation and inversion should
enable integration of solid-state, complementary logic elements on
an individual CNFET.\cite{derycke}

Low bias $I-V_g$ data (Fig.~2(a)) shows that the nanotube band gap
corresponds to $\delta V_g = 2.0$V. Based on the nanotube
diameter, measured by AFM, we estimate the band gap to be $E_G =
0.6$eV. The ratio $\alpha_g = E_G/\delta V_g = C_g/C$ is the
electrostatic ``lever arm'' of the gate (here $C_g$ and $C$ are
the gate capacitance and total capacitance of the CNFET,
respectively). Our measured value $\alpha_g = 0.33$ is an order of
magnitude larger than that previously observed in CNFET
devices.\cite{mceuen_disorder} We confirm this measurement of
$\alpha_g$ with Coulomb blockade (CB) data\cite{nato_marcus} from
which $C_g/C$ can be calculated. Figure 2(d) shows the
differential conductance of the {\it same} CNFET as a function of
the bias and gate voltages at $T = 5$K in the hole accumulation
regime. Current through the device is blocked within the black
diamonds in the $V-V_g$ plane. Extraordinarily regular size and
spacing of Coulomb diamonds indicates that the CNFET forms a
single quantum dot, a signature of {\it ballistic
transport}\cite{bock_set} within the semiconducting nanotube on
the scale of 800nm at sufficient electrostatic doping. This is in
contrast to early proposals of diffusive transport in
CNFETs.\cite{martel_fet,mceuen_disorder} We find the capacitive
lever arm of the gate by comparing the charging energy of the
nanotube to the period of Coulomb oscillations: $\alpha_g =
$~6mV/15mV = 0.4, close to the value derived from room temperature
data. We consistently find gate lever arms of $0.2-0.4$, strong
evidence that annealing of the oxide film in hydrogen and the
associated reduction in trap density improves gating action and
ambipolar transport in nanotube devices. Extremely good
transconductance and subthreshold swing values compare favorably
to ambipolar CNFETs fabricated using two orders of magnitude
thinner dielectric layers\cite{bachtold_logic} or electrolytic
gates\cite{kruger}.

We quantitatively explain the enhanced gate coupling as due to
improvement of the gate oxide layer through the hydrogen anneal.
Silicon dioxide has interface and bulk (dangling bond) traps whose
charge state changes with gate voltage (Fig.~2(c)). High quality
as-grown oxide films have trap densities $D_{it} \sim 5\! \times
\! 10^{11}$/cm$^2$-eV and $N_{ot}\sim 5\! \times \!
10^{11}$/cm$^2$ for interface traps at midgap and oxide traps,
respectively.\cite{wolf} Interface traps are populated
continuously as the gate voltage is tuned, while oxide traps are
charged only with injection at gate fields above $3\! \times \!
10^5$V/cm. Here we focus on interface traps; we discuss bulk oxide
traps in the context of large gate fields in the next paragraph.
The backgate voltage is screened effectively by traps within a
$L\! \times\! (2\pi h)$ rectangular cutout of the outer cylinder
in the coaxial cable approximation for CNFET capacitance ($L$ is
the device length, $h$ the dielectric
thickness\cite{capacitance_foot}). This implies that in the area
beneath the CNFET there are 7000/eV-$\mu$m interface traps, much
larger than the total device capacitance, $C = 130e$/V-$\mu$m. The
lever arm is approximately the ratio between the self capacitance
and the trap density: $\alpha_g \sim 130/7000 = 0.018$, a value
similar to that observed by other groups.\cite{bock_set} Our
measurement $\alpha_g = 0.4$ indicates that the interface trap
density is reduced by a factor of 50 to $D_{it} \sim 2\! \times \!
10^{10}$/cm$^2$-eV, which is consistent with expectations for the
anneal conditions.\cite{wolf}

Figure 3(a)-(c) compares room temperature $I-V_g$ curves of an
ambipolar, Co-contacted CNFET for different $V_g$ ranges. There is
reproducible hysteresis in the $I-V_g$ curves that becomes larger
as the range of $V_g$ is increased, indicating that it originates
from avalanche injection into bulk oxide traps. Hysteresis can
become so large that the device varies between depletion mode
(normally-``on'' at $V_g=0$) and enhancement mode
(normally-``off'') behavior. We determine the location and sign of
trapped charges by comparing the direction of the hysteresis to
the avalanching field. After a sweep to positive $V_g$, the CNFET
threshold voltage moves toward more {\it positive} gate values
indicating injection of {\it negative} charges into oxide traps.
Traps are populated by electrons injected from the CNFET channel,
where the electric field is highest due to the cylindrical device
geometry. These charged traps are near the device\cite{where},
perhaps the same scattering sites imaged in SGM (Fig.~1(c) and
(e)).

The hysteresis allows the device to function as a non-volatile memory
cell, similar in operation to electrically erasable, programmable
read-only memories (EEPROMs\cite{horowitz}). To read out the memory, a
1M$\Omega$ load resistor is added to create a voltage divider circuit
(Fig.~3(e), inset). Read ($V_{in} = 0$) and write ($V_{in} = +20$V or
-20V) are applied to the input (backgate) terminal. Logical ``1'' (``0'')
is defined as $V_{out} = 1$V (0V). To write a ``1'' (``0'') to the memory
cell, $V_{in}$ is switched rapidly to -20V (+20V) and back to 0, so the
CNFET is ``on'' (``off'') at the read voltage (c.f.~Fig.~3(c)). Figure
3(e) shows the memory cell output as a function of time while a series of
data bits is written. Irregular sequence of bit values and writing times
is used to demonstrate stability of the memory cell. The CNFET-based
memory is non-volatile at room temperature, with bit storage times of at
least 16 hours. The device read/write speed is limited by trap charging
times, which are much shorter than the 100Hz pulses used here. The number
of occupied trap charges is estimated from the threshold voltage shift and
an effective capacitance: $\Delta Q_{ot} = C_{\rm eff} \Delta V_T$, where
$C_{\rm eff}$ is related to the active oxide region for data storage,
located near the Schottky barriers (of order $20$nm
width\cite{jacques_gaps}). The bit is stored in no more than 2$e$, 70$e$
and 200$e$ charges (for Fig.~3(a)-(c), respectively), quantitatively
similar to floating gate single-electron memory devices based on
conventional silicon transistors.\cite{chou_singelmem} More controllable,
single-electron data storage may be achieved by intentional fabrication of
charge storage sites in the active device region.

In conclusion, air-stable and intrinsic n-type, ambipolar
CNFETs\cite{martel_prl} can be fabricated by controlling the
Schottky barrier\cite{freitag_apl}. Excellent switching
characteristics are achieved when the gate oxide is subject to
hydrogen anneal. Our observation of an order of magnitude
improvement in gate lever arm is directly verified using Coulomb
blockade data in the strong inversion regime. The existence of a
single quantum dot within the nanotube at low temperatures
suggests that at high doping levels semiconducting nanotubes, like
their metallic counterparts, are ballistic conductors on the
1$\mu$m scale. Finally, we have harnessed charge injection into
the oxide at large gate voltage to construct a nanotube-based
molecular memory cell with data storage nearly at the
single-electron level.\cite{fuhrer}

We acknowledge helpful conversations with C. Kane and E. Mele. This work
is supported by the LRSM (a NSF MRSEC DMR00-79909 (ATJ, MR, MF)), and the
LRSM Research Experience for Undergraduates program (KVT).

\begin{figure}
\begin{center}
\epsfig{figure=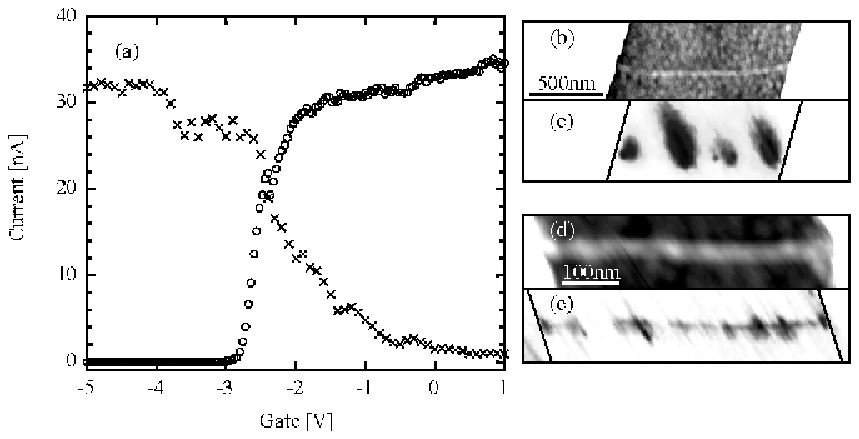}
\end{center}
\caption{Complementary CNFETs: (a) Ambient $I-V_g$ curves for
Cr/Au-contacted p-type CNFET (crosses) and Co-contacted n-type CNFET
(circles) (bias voltage $V_{ds} = 10$mV). ``On''-state resistance is
300k$\Omega$ and switching ratio exceeds $10^5$. (b)-(c) Topographic AFM
and SGM scans for a CNFET with Cr/Au electrodes. SGM ($V_t = +2$V,
$V_{ds}=50$mV) shows a series of well-localized, gateable regions along
the CNFET. Grayscale in (c) represents transport currents between 0
(black) and 400nA (white). (d)-(e) Corresponding data for a Co-contacted
CNFET in the ambient. SGM ($V_t = -1.2$V, $V_{ds}=200$mV) shows
complementary gateable regions sensitive to the negative tip voltage. The
grayscale in (e): 0 (black) to 250nA (white).}
\end{figure}

\begin{figure}
\begin{center}
\epsfig{figure=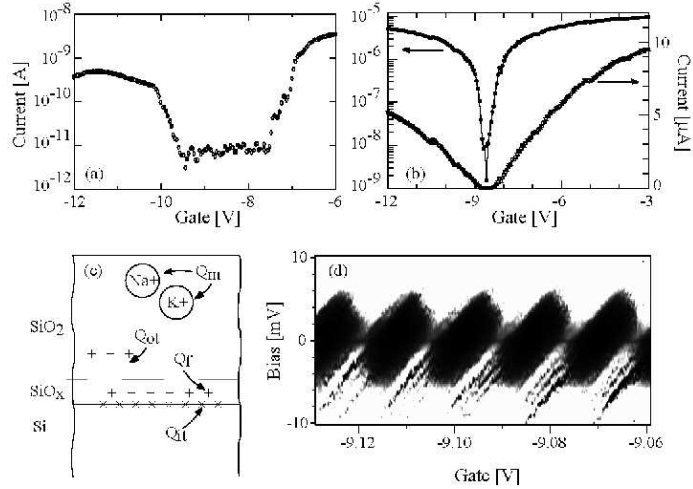}
\end{center}
\caption{(a) $I-V_g$ characteristic ($V_{ds}=0.5$mV) shows ambipolar
transistor action. (b) Same sample at $V_{ds}=1$V shows transconductance
of approximately $3\mu$A/V for electrons and $2\mu$A/V for holes and
subthreshold swing of 100mV/decade. (c) Schematic depicts interface
($Q_{it}$) and bulk ($Q_{ot}$) oxide traps. (d) Coulomb Blockade data for
the CNFET in hole conduction regime. Uniform CB is consistent with a long
mean free path in doped semiconducting nanotubes.}
\end{figure}

\begin{figure}
\begin{center}
\epsfig{figure=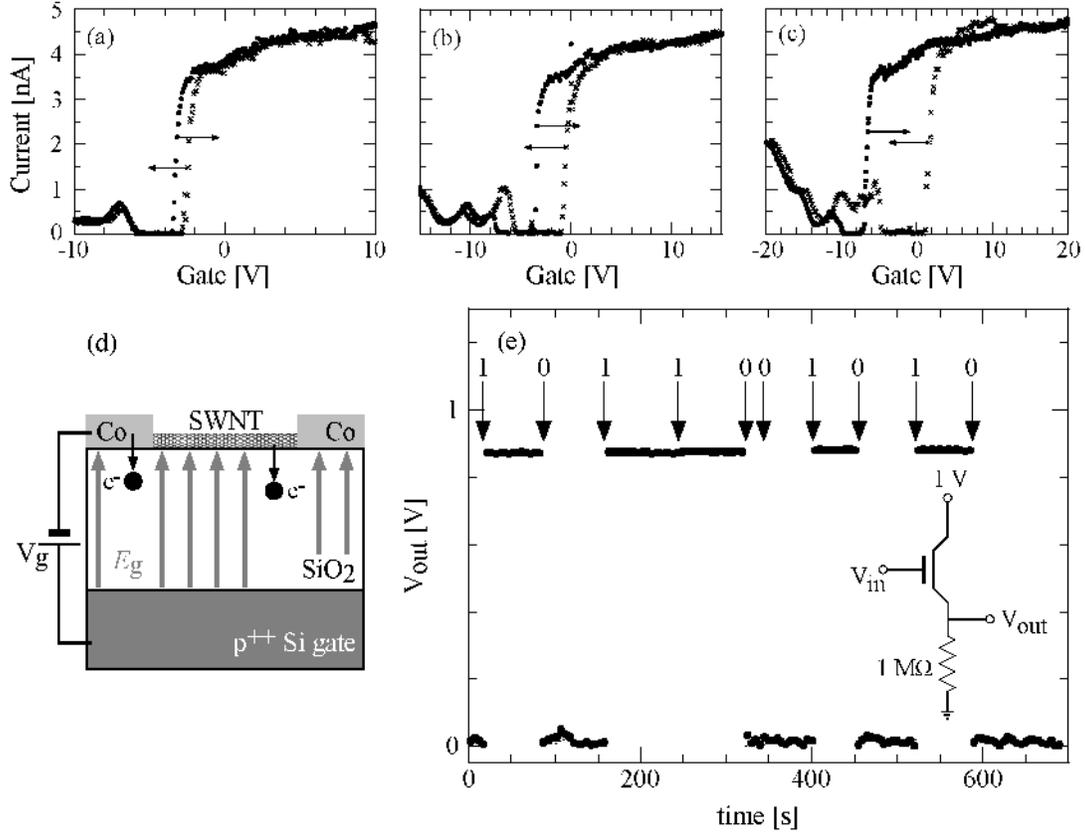}
\end{center}
\caption{(a)-(c) High vacuum $I-V_g$ data at $V_{ds}=0.5$mV. Device
hysteresis increases steadily with increasing $V_g$ due to avalanche
charge injection into bulk oxide traps. (d) Diagram of avalanche injection
of electrons into bulk oxide traps from the CNFET channel. (e) Data from
CNFET-based non-volatile molecular memory cell. A series of bits is
written into the cell (see text) and the cell contents are continuously
monitored as a voltage signal ($V_{out}$) in the circuit shown in the
inset.}
\end{figure}

\begin{figure}
\begin{center}
\epsfig{figure=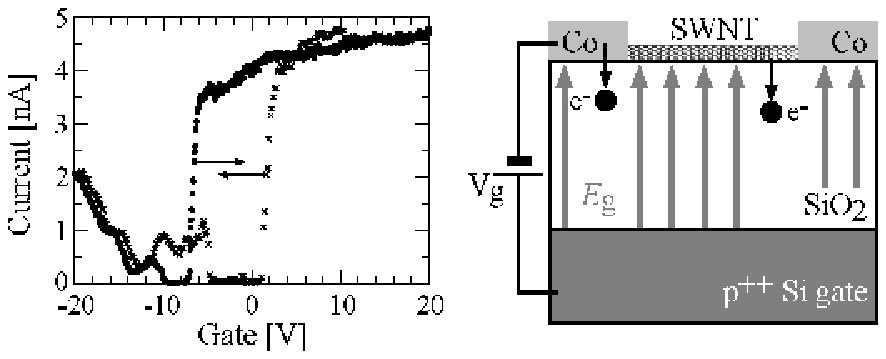}
\end{center}
\caption{Table of Contents graphic}
\end{figure}

\begin{references}

\bibitem{tans_fet} S.~J. Tans, R.~M. Verschueren, and C. Dekker, Nature
{\bf 393}, 49
(1998).

\bibitem{martel_fet} R. Martel, T. Schmidt, T. Hertel, and P. Avouris,
Appl. Phys.
Lett. {\bf 73}, 2447 (1998).

\bibitem{collins_oxygen} P.~G. Collins, K. Bradley, M. Ishigami, and A.
Zettl, Science {\bf
287}, 1801 (2000).

\bibitem{freitag_apl} M. Freitag {\it et~al.}, Appl. Phys. Lett. {\bf 79},
3326 (2001).

\bibitem{martel_prl} R. Martel {\it et al.}, Phys. Rev. Lett. {\bf 87},
256805 (2001).

\bibitem{derycke} V. Derycke, R. Martel, J. Appenzeller, and Ph. Avouris,
Nano Lett.
{\bf 1}, 453 (2001).

\bibitem{derycke_apl} V. Derycke, R. Martel, J. Appenzeller, and Ph.
Avouris, Appl.
Phys. Lett. {\bf 80}, 2773 (2002).

\bibitem{bachtold_logic} A. Bachtold, P. Hadley, T. Nakanishi, and C.
Dekker, Science {\bf
294}, 1317 (2001).

\bibitem{liu_logic} X. Liu, C. Lee, C. Zhou, and J. Han, Appl. Phys. Lett.
{\bf 79},
3329 (2001).

\bibitem{bock_Kdop} M. Bockrath {\it et~al.}, Phys. Rev. B {\bf 61},
R10606 (2000).

\bibitem{dai_nature} J. Kong {\it et~al.}, Nature {\bf 395}, 878 (1998).

\bibitem{hafner_nature} J.~H. Hafner, C.~L. Cheung, and C.~M. Lieber,
Nature {\bf 398},
761 (1999).

\bibitem{snyder} C.~E. Snyder {\it et~al.}, Int. Pat. WO 9/07163 (1989).

\bibitem{wolf} S. Wolf and R.~N. Tauber, {\em Silicon Processing for the
VLSI
Era} (Lattice Press, Sunset Beach, CA, 1986).

\bibitem{tans_bach_sgm} S.~J. Tans and C. Dekker, Nature {\bf 404}, 834
(2000); A.
Bachtold {\it et~al.}, Phys. Rev. Lett. {\bf 84}, 6082 (2000).

\bibitem{freitag_ropes} M. Freitag, M. Radosavljevi\'{c}, W. Clauss, and
A.~T. Johnson,
Phys. Rev. B {\bf 62}, R2307 (2000).

\bibitem{crc} {\em CRC Handbook of Chemistry and Physics}, 82nd ed. (CRC
Press,
Cleveland, OH, 2001).

\bibitem{martel_iedm} R. Martel, H.-S.~P. Wong, K. Chan, and Ph. Avouris,
IEDM, 159
(2001).

\bibitem{marko_unpub} M. Radosavljevi\'{c}, Ph. D. Dissertation,
University of
Pennsylvania (2001).

\bibitem{mceuen_pndot} J. Park and P.~L. McEuen, Appl. Phys. Lett. {\bf
79}, 1363 (2001).

\bibitem{sze} S.~M. Sze, {\em Semiconductor Devices, Physics and
Technology}
(John Wiley and Sons, New York, 1985).

\bibitem{kruger} M. Kr\"{u}ger {\it et~al.}, Appl. Phys. Lett. {\bf 78},
1291
(2001).

\bibitem{mceuen_disorder} P.~L. McEuen {\it et~al.}, Phys. Rev. Lett. {\bf
83}, 5098 (1999).

\bibitem{nato_marcus} L.~P. Kouwenhoven, C.~M. Marcus, P.~L. McEuen, S.
Tarucha, R.~M.
Westervelt, and N.~S. Wingreen, in {\em Mesoscopic Electron
Transport} (Plenum, New York, 1997), pp.~$105$-$214$.

\bibitem{bock_set} M. Bockrath {\it et~al.}, Science {\bf 275}, 1922
(1997).

\bibitem{capacitance_foot} CNFET capacitance per unit length is derived
from the cylindrical
geometry: $C/L = 2 \pi \varepsilon \varepsilon_o/\ln (2h/r)$ or
21aF/$\mu$m = 130$e/\mu$m-V for our devices.

\bibitem{where} The effect of charge trapping in the impurities which are
deposited on the surface during processing is minimized by
measuring in $10^{-6}$~Torr vacuum.

\bibitem{horowitz} P. Horowitz and W. Hill, {\em The Art of Electronics}
(Cambridge
University Press, New York, NY, 1980).

\bibitem{jacques_gaps} J. Lefebvre, M. Radosavljevi\'{c}, and A.~T.
Johnson, Appl. Phys.
Lett. {\bf 76}, 3828 (2000).

\bibitem{chou_singelmem} L. Guo, E. Leobandung, and S.~Y. Chou, Science
{\bf 274}, 2069 (1997).

\bibitem{fuhrer} During preparation of this article, we became aware of
similar
research done by M. S. Fuhrer and collaborators. They have
observed an even higher ``effective'' mobility than we have, in
agreement with our claim of near-ballistic transport in these
devices. Additionally, they have observed the memory effect we
investigated.

\end{references}
\end{document}